\documentstyle[preprint,aps,pre]{revtex}
\begin{document}
\tighten
\title{Transport coefficients of a heated granular gas}
\author{Vicente Garz\'{o}\footnote[1]{Electronic address: vicenteg@unex.es}}
\address{Departamento de F\'{\i}sica, Universidad de Extremadura, E-06071
Badajoz, Spain}
\author{Jos\'e Mar\'{\i}a Montanero\footnote[2]
{Electronic address: jmm@unex.es}}
\address{Departamento de Electr\'onica e Ingenier\'{\i}a Electromec\'anica,
Universidad de Extremadura, E-06071 Badajoz, Spain}
\date{\today}
\maketitle

\begin{abstract}

The Navier-Stokes transport coefficients of a granular gas
are obtained from the Chapman-Enskog solution to the Boltzmann equation.
The granular gas is heated by the action of an
external driving force (thermostat) which does work
to compensate for the collisional loss of energy. Two types of thermostats 
are considered: (a) a deterministic force proportional to the 
particle velocity (Gaussian thermostat), and (b) a random external force
(stochastic thermostat). As happens in the free cooling case,
the transport coefficients are determined from
linear integral equations which can be approximately solved by means of a 
Sonine polynomial expansion. In the leading order, we get those coefficients 
as explicit functions of the restitution coefficient $\alpha$. The results 
are compared with those obtained in the free cooling case,
indicating that the above thermostat
forces do not play a neutral role in the transport.
The kinetic theory results are also compared  with those obtained 
from  Monte Carlo simulations of the Boltzmann equation for
the shear viscosity.
The comparison shows an excellent agreement between theory
and simulation over
a wide range of values of the restitution coefficient.
Finally, the expressions of the transport coefficients for a gas of
inelastic hard spheres are extended to the revised Enskog theory for a
description at higher densities.

\end{abstract}

\draft

\bigskip \narrowtext

\section{Introduction}
\label{sec1}

The usefulness of fluid-like type equations to describe systems of granular 
particles in rapid, dilute flow has been recognized for many years. The 
essential difference from ordinary fluids is the absence of energy 
conservation, yielding subtle modifications of the conventional 
Navier-Stokes equations for states with small spatial gradients of the 
hydrodynamic fields. Although many efforts have been made in the past few 
years in the understanding of these systems,  the analysis of the 
influence of dissipation on the 
transport coefficients still remains a topic of interest and controversy. 
For a low density gas, these coefficients may be determined from 
the Boltzmann equation modified to account for inelastic binary collisions. 
The idea is to extend the Chapman-Enskog method \cite{FK72} to the inelastic 
case by expanding around the local version of the homogeneous cooling state 
(HCS), i.e., a reference state in which all the time dependence occurs 
through the granular temperature. In the first-order of the expansion, explicit 
expressions for the transport coefficients as functions of the restitution 
coefficient have been obtained in the case of hard spheres \cite{BDKS98} as 
well as for a $d$-dimensional system\cite{BC01}. This analysis has been 
also extended to higher densities in the context of the Enskog equation 
\cite{GD99}. The results obtained in the latter theory describe very well 
the hydrodynamic profiles obtained in a recent 
experimental study of a three-dimensional system of mustard seeds fluidized 
by vertical vibrations of the container \cite{YHCMW02}.

One of the main difficulties in obtaining the above transport coefficients 
lies in the fact that, in contrast to what happens for 
molecular fluids, the reference state (zeroth-order solution of the 
Chapman-Enskog expansion) depends on time due to the dissipation of energy 
through collisions. In addition, it is also well-known that this state is 
unstable to long enough wavelength perturbations so that the state becomes 
inhomogeneous for long times\cite{D01}.
To overcome such difficulties, one possibility 
is to introduce external forces to accelerate the particles and hence 
compensate for collisional cooling. As a consequence, the corresponding 
reference state is stationary and linearly stable against spatial
inhomogeneities. This mechanism of energy
input (different from those in shear flows or flows through vertical pipes) 
has been used by many authors \cite{WM96,varios,NE98,NETP99,PTNE02,MS00}
in the past years to analyze different problems, such as
non-Gaussian properties (cumulants, high energy tails) of the velocity 
distribution function \cite{NE98,MS00}, long-range correlations \cite{NETP99}, and 
collisional statistics and short scale structure \cite{PTNE02}. Since the latter 
requires the solution of the corresponding linearized hydrodynamic equations around 
the homogeneous state, the explicit expressions for the transport 
coefficients are needed. Given that the dependence of these coefficients on 
the restitution coefficient is not known, the expressions
of these coefficients are usually assumed to be the same as those for the                   
{\em elastic} gas. However, according to the results derived in the 
free cooling case \cite{BDKS98,BC01,GD99}, the above assumption could be only justified in 
the {\em small} inelasticity limit.

The goal of this paper is to determine the transport coefficients of a 
heated granular gas. This allows us to measure the new effects induced by 
the external force on transport by comparison with the results derived in 
the unforced case\cite{BDKS98,BC01}. In addition, one could also assess to what 
extent the previous results on short and large structure\cite{NETP99,PTNE02} 
are indicative of what happens for finite degree of dissipation. 
There are different 
mechanisms to inject energy to the gas. Here, the fluidization is driven by 
the action of external forces (thermostats) acting locally on each particle. 
In this paper we will consider two types of thermostats: 
the Gaussian and the stochastic thermostats. In 
the case of the Gaussian thermostat, the gas is heated by the action of an 
external force proportional to the peculiar velocity. This type of 
``anti-drag'' force can be justified by Gauss's principle of least 
constraints \cite{EM90} and has been widely used in nonequilibrium molecular 
dynamics simulations of molecular fluids. Another mechanism for 
thermostatting the system is to assume that the particles are subjected to a 
random external force, which gives frequent kicks to each particle between 
collisions. If this stochastic force has the properties
of a white noise, it gives rise to a Fokker-Planck diffusion term in the 
Boltzmann equation \cite{NE98}.

It must be remarked that in most experiments, energy is added to the
granular gas through a boundary, causing gradients in the energy 
perpendicular to that boundary. In this sense, we do not claim that the above
forcing terms are the most suited to model any particular real system.
However, they have the advantage of that it can be incorporated into
the kinetic theory very easily. This allows for instance, to test the
assumptions of the Chapman-Enskog method through a direct comparison
with computer simulations, such as Monte Carlo or molecular dynamics
simulations\cite{BSSS99}. 

The plan of the paper is as follows. In Sec.\ \ref{sec2}, we review the 
Boltzmann equation and associated macroscopic conservation laws in the 
presence of the external forces discussed above. The Chapman-Enskog method 
for solving this equation is presented in Sec.\ \ref{sec3} and subsequently 
applied to the cases of Gaussian and stochastic thermostats. As happens in 
the free cooling case\cite{BDKS98}, the transport
coefficients are determined from linear integral equations which
can be approximately solved by means of a Sonine polynomial
expansion. In the leading order, we get the transport coefficients as
explicit functions of the restitution coefficient.
Section \ref{sec3} ends with
a comparison between the results derived here in both driven cases with 
those previously obtained in the unforced case \cite{BC01}. Such a comparison 
shows that, in general, the thermostats do not play a neutral role in the 
transport since they clearly affect the dependence of the Navier-Stokes 
transport coefficients on the dissipation. In order to check the 
degree of reliability of the Sonine approximation, a comparison with direct 
Monte Carlo simulation of the Boltzmann equation is carried out in Sec.\ 
\ref{sec4}. More specifically, the simulations are performed for a gas 
undergoing uniform shear flow, using the Gaussian and the stochastic 
thermostats to control 
inelastic cooling. In the long time limit, a (reduced) shear 
viscosity can be measured in both simulations. 
The comparison with the Chapman-Enskog solution shows an excellent 
agreement, indicating that the Sonine results have an accuracy comparable to 
that for elastic collisions. The paper is closed in
Sec.\ \ref{sec5} with a
brief summary and discussion of the results presented. In addition, the
corresponding expressions of the transport coefficients for a granular gas
of hard spheres in the framework of Enskog kinetic theory are also
displayed in Appendix \ref{appB}.

\section{Heated granular gases}
\label{sec2}

We consider a granular gas composed by smooth inelastic disks ($d=2$)
or spheres ($d=3$) of mass $m$ and diameter $\sigma$.
The inelasticity of collisions among all pairs
is characterized by a constant restitution coefficient $\alpha\leq 1$.
In the low-density regime, the evolution
of the one-particle velocity distribution function $f({\bf r},{\bf v};t)$ is 
given by the Boltzmann kinetic equation \cite{GS95,BDS97} 
\begin{equation}
\label{2.1}
\left(\partial _{t}+{\bf v}_{1}\cdot \nabla +{\cal F}\right) f({\bf r},{\bf 
v}_{1},t)=J\left[{\bf v}_{1}|f(t),f(t)\right] \;,
\end{equation}
where the Boltzmann collision operator $J[{\bf v}_1|f,f]$ is 
\begin{eqnarray}
\label{2.2}
J\left[{\bf v}_{1}|f,f\right]  &=&\sigma^{d-1}\int d{\bf v}_{2}\int 
d\widehat{\bbox {\sigma }}\,\Theta (\widehat{\bbox {\sigma}}\cdot {\bf 
g})(\widehat{\bbox {\sigma }}\cdot {\bf g})  \nonumber \\
&&\times \left[ \alpha^{-2}f({\bf r},{\bf v}_{1}^{\prime },t)f(
{\bf r},{\bf v}_{2}^{\prime },t)-f({\bf r},{\bf v}_{1},t)f({\bf r},
{\bf v}_{2},t)\right] \;. 
\end{eqnarray}
In Eq.\ (\ref{2.1}), ${\cal F}$ is an operator representing
the effect of an external forcing which injects energy into the granular gas
allowing it to reach a steady state. Furthermore, $d$ is the dimensionality 
of the system, $\widehat{\bbox {\sigma}}$ is a unit vector along their line 
of centers, $\Theta $ is
the Heaviside step function, and ${\bf g}={\bf v}_{1}-{\bf v}_{2}$. The
primes on the velocities denote the initial values $\{{\bf v}_{1}^{\prime },
{\bf v}_{2}^{\prime }\}$ that lead to $\{{\bf v}_{1},{\bf v}_{2}\}$
following a binary collision: 
\begin{equation}
\label{2.3}
{\bf v}_{1}^{\prime}={\bf v}_{1}-\case{1}{2}\left(1+\alpha^{-1}\right)
(\widehat{\bbox {\sigma}}\cdot {\bf g})\widehat{\bbox {\sigma}},
\quad {\bf v}_{2}^{\prime }={\bf v}_{2}+\case{1}{2}\left( 
1+\alpha^{-1}\right) (\widehat{\bbox {\sigma}}\cdot 
{\bf g})\widehat{\bbox{\sigma}}
\end{equation}

The macroscopic balance equations for density $n$, momentum $m{\bf u}$, and 
energy $\case{d}{2}nT$ follow directly from Eq.\ ({\ref{2.1}) by multiplying 
with $1$, $m{\bf v}_1$, and $\case{m}{2}v_1^2$ and integrating over ${\bf 
v}_1$:
\begin{equation}
\label{2.4}
D_{t}n+n\nabla \cdot {\bf u}=0\;,  
\end{equation}
\begin{equation}
\label{2.5}
D_{t}u_i+(mn)^{-1}\nabla_j P_{ij}=0\;,  
\end{equation}
\begin{equation}
\label{2.6}
D_{t}T+\frac{2}{dn}\left(\nabla \cdot {\bf q}+P_{ij}\nabla_j u_i\right) 
=-(\zeta-\Lambda)T\;.  
\end{equation}
In the above equations, $D_{t}=\partial _{t}+{\bf u}\cdot \nabla$ is the
material derivative, 
\begin{equation}
{\sf P}=\int d{\bf v}\,m{\bf V}{\bf V}\,f({\bf v})
 \label{2.7}
\end{equation}
is the pressure tensor, 
\begin{equation}
{\bf q}=\int d{\bf v}\,\case{1}{2}m V^{2}{\bf V}\,
f({\bf v})
\label{2.8}
\end{equation}
is the total heat flux, and ${\bf V}={\bf v}-{\bf u}$ is the peculiar
velocity. In the right hand side of the temperature equation (\ref{2.6}), 
the cooling rate $\zeta$ (measuring the rate of energy loss due to 
dissipation) and the source term $\Lambda$ (measuring the rate of heating 
due to the external force) are given by  
\begin{equation}
\label{2.9}
\zeta=-\frac{1}{dnT}\int\, d{\bf v} mv^2J[f,f],
\end{equation}
\begin{equation}
\label{2.10}
\Lambda=-\frac{1}{dnT}\int\, d{\bf v} mv^2{\cal F}f({\bf v}).
\end{equation}
It is assumed that the external driving does not change the
number of particles or the momentum, i.e.,
\begin{equation}
\label{2.10bis}
\int d{\bf v}\, {\cal F} f({\bf v})=\int d{\bf v}\, {\bf v}
{\cal F} f({\bf v})=0.
\end{equation}

In the case of elastic particles ($\alpha=1$) and in the absence of external
forcing (${\cal F}=0$), it is well known that the long-time {\em 
uniform} solution of Eq.\ (\ref{2.1}) is the Maxwell-Boltzmann distribution 
function. However, if the particles collide inelastically ($\alpha<1$) and 
${\cal F}=0$, a steady state is not possible in uniform situations 
since the temperature decreases monotonically in time. In this case,
Goldshtein and Shapiro \cite{GS95} showed that Eq.\ (\ref{2.1}) admits
an isotropic solution, describing the homogeneous cooling state
(HCS), in which all the time dependence of $f$ occurs only
through the thermal velocity
$v_0(t)=\sqrt{2T(t)/m}$: $f({\bf v},t)\rightarrow n
\pi^{-d/2}v_0^{-d}(t) \Phi({\bf v}/v_0(t))$. So far, the exact form of 
$\Phi$ has not been found, although a good approximation for thermal
velocities can be
obtained from an expansion in Sonine polynomials. In the leading order, 
$\Phi$ is given by 
\begin{equation}
\label{2.11}
\Phi({\bf v}^*)\rightarrow \left\{
1+\frac{c}{4}\left[v^{*4}-(d+2)v^{*2}+
\frac{d(d+2)}{4}\right]\right\}e^{-v^{*2}},\quad {\bf v}^*={\bf v}/v_0,
\end{equation}
where the estimated value of $c$ is \cite{NE98}
\begin{equation}
\label{2.12}
c=\frac{32(1-\alpha)(1-2\alpha^2)}{9+24d-\alpha(41-8d)+30(1-\alpha)\alpha^2}.
\end{equation}
The estimate (\ref{2.12}) presents a quite good agreement with Monte Carlo 
simulations of the Boltzmann equation \cite{MS00,BRC96}.

However, by driving a granular gas by boundaries or external fields it can 
reach a steady state. The energy injected in the gas may exactly 
compensate for the energy dissipated by collisional cooling.
The same effect can be obtained by means of external forces
acting locally on each particle.
These forces, which we will call {\em thermostats}, are represented by the
operator ${\cal F}$ in Eq.\ (\ref{2.1}) and depend on the state of the system.
Several types of thermostats can be used. Here, we will
consider two. One of them is a deterministic thermostat widely used in 
nonequilibrium molecular dynamics simulations of elastic particles\cite{EM90},
which is based on Gauss's principle of least constraints. In this
case, ${\cal F}$ is given by \cite{MS00}
\begin{equation} 
\label{2.13}
{\cal F}f({\bf v})=\frac{1}{2}\zeta\frac{\partial}{\partial {\bf v}}\cdot 
\left[{\bf v}f({\bf v})\right],
\end{equation}
where, according to Eqs.\ (\ref{2.6}) and (\ref{2.10}), the thermostat has
been adjusted to get a constant temperature in the long time limit. It must 
be pointed out that the corresponding Boltzmann equation (\ref{2.1}) for 
this Gaussian thermostat force is formally identical with the Boltzmann 
equation in the HCS (i.e. with ${\cal F}=0$) when both 
equations are written in terms of the reduced distribution $\Phi(v^*)$. As a 
consequence, the result (\ref{2.12}) applies to this thermostatted case as 
well.

Another way of heating the gas is by means of a stochastic force assumed to 
have the form of a Gaussian white noise \cite{WM96}. The corresponding 
operator ${\cal F}$ has a Fokker-Planck form \cite{NE98}
\begin{equation}
\label{2.14}
{\cal F}f({\bf v})=-\frac{1}{2}\frac{T}{m}\zeta \left(
\frac{\partial}{\partial {\bf v}}\right)^2f({\bf v}),
\end{equation}
where again the strength of the correlation has been chosen to achieve 
a time independent temperature.  
By using this thermostat, van Noije and Ernst \cite{NE98} have studied the 
stationary solution to the uniform Boltzmann equation (\ref{2.1}) and found 
for the coefficient $c$ the value
\begin{equation}
\label{2.15}
c=\frac{32(1-\alpha)(1-2\alpha^2)}{73+56d-3\alpha(35+8d)+30(1-\alpha)\alpha^2}.
\end{equation} 
As happened in the Gaussian case, recent Monte Carlo simulations of the 
Boltzmann equation \cite{MS00} agree quite well with the Sonine estimate 
(\ref{2.15}).

\section{Chapman-Enskog solution: Navier-Stokes transport coefficients}
\label{sec3}

As said in the Introduction, our goal is to get the Navier-Stokes transport 
coefficients in the presence of the thermostats introduced in the previous 
Section. This allows one to assess the influence of these thermostats on 
transport by comparison with the results
obtained in the free cooling case\cite{BDKS98,BC01}.
To do that, we consider now a spatially {\em 
inhomogeneous} state, created by initial preparation or by boundary 
conditions. We assume that the spatial variations of $n$, ${\bf u}$, and $T$ 
are small on the scale of the mean free path. Under these conditions, the 
Chapman-Enskog method \cite{FK72} provides a solution to the Boltzmann 
equation based on an expansion around the {\em local} version of the heated 
homogeneous state induced by the thermostat forces. 
This is obtained from 
the above states by replacing the temperature, density, and flow velocity by 
their nonequilibrium local values. As a consequence, the local version of the 
operators ${\cal F}$ consists of replacing ${\bf v}\rightarrow {\bf V}={\bf 
v}-{\bf u}$ in Eqs.\ (\ref{2.13}) and (\ref{2.14}). 
 
The Chapman-Enskog method assumes the existence of a {\em normal} solution 
in which all the space and time dependence of the distribution function appears 
through a functional dependence on the hydrodynamic fields
\begin{equation}
\label{3.1}
f({\bf r},{\bf v},t)=f\left[{\bf v}|n(t), {\bf u}(t), T(t)\right].
\end{equation}
For small spatial variations, this functional dependence can be made local 
in space and time through an expansion in gradients of the fields.  
To generate the expansion, it is convenient to write $f$
as a series expansion in a
formal parameter $\epsilon $ measuring the nonuniformity of the system, 
\begin{equation}
f=f^{(0)}+\epsilon \,f^{(1)}+\epsilon^2 \,f^{(2)}+\cdots \;,
\label{3.2}
\end{equation}
where each factor of $\epsilon$ means an implicit gradient of a
hydrodynamic field. The local reference state $f^{(0)}$ is chosen such
that it has the same first moments as the 
exact distribution $f$, or
equivalently, the remainder of the expansion must obey the orthogonality
conditions 
\begin{equation}
\int d{\bf v}\left[ f({\bf v})-f^{(0)}({\bf v})\right]
=0\;,\quad \int d{\bf v} {\bf v}\left[ f({\bf v})-f^{(0)}({\bf 
v})\right] ={\bf 0}\;,  \label{3.3}
\end{equation}
\begin{equation}
\int d{\bf v}v^{2}\left[ f({\bf v})-f^{(0)}({\bf 
v})\right] =0\;.  \label{3.4}
\end{equation}
The time derivatives of the fields are also expanded as $\partial
_{t}=\partial _{t}^{(0)}+\epsilon \partial _{t}^{(1)}+\cdots $. The
coefficients of the time derivative expansion are identified from the
balance equations (\ref{2.4})--(\ref{2.6}) with a representation of the
fluxes, the cooling rate $\zeta$ and the heating term $\Lambda$ in the 
macroscopic balance equations as a
similar series through their definitions as functionals of the distribution 
$f$. This is the usual Chapman-Enskog method for solving kinetic
equations \cite{FK72}. The main difference with respect to the unforced 
case \cite{BDKS98} is that now (as happens in the elastic case) the 
sink term in the energy equation is zero, so that the terms coming from the time
derivative $\partial_{t}^{(0)}$ vanish.

Now, we derive the corresponding hydrodynamic equations in the 
presence of the Gaussian and the stochastic thermostats.

\subsection{Gaussian thermostat}

In the case of the Gaussian thermostat (\ref{2.13}), the Boltzmann equation 
becomes 
\begin{equation}
\label{3.5}
\partial _{t}f+{\bf v}\cdot \nabla f+\frac{1}{2}\zeta
\frac{\partial}{\partial {\bf V}}\cdot\left({\bf V}f\right)
=J\left[f,f\right] \;.
\end{equation}
Substitution of the Chapman-Enskog solution (\ref{3.2}) into Eq.\ 
(\ref{3.5}) leads to different kinetic equations for the distributions 
$f^{(k)}$. To zeroth order in $\epsilon$, the Boltzmann equation reads
\begin{equation}
\label{3.6}
\frac{1}{2}\zeta^{(0)}
\frac{\partial}{\partial {\bf V}}\cdot \left({\bf V}f^{(0)}\right)
=J\left[f^{(0)},f^{(0)}\right],  
\end{equation} 
where use has been made of the macroscopic balance equations at this order
\begin{equation}
\label{3.7}
\partial_t^{(0)}n=0,\quad \partial_t^{(0)}{\bf u}=0, \quad 
\partial_t^{(0)}T=0.
\end{equation}
Here, the cooling rate $\zeta^{(0)}$ is determined by Eq.\ (\ref{2.9}) to 
zeroth order
\begin{equation}
\label{3.8}
\zeta^{(0)}=-\frac{1}{dnT}\int\, d{\bf v} mv^2J[f^{(0)},f^{(0)}].
\end{equation}
The solution to Eq.\ (\ref{3.6}) $f^{(0)}=f^{(0)}(V)$ is isotropic so that 
the zeroth order approximations to the pressure tensor and heat flux are
\begin{equation}
\label{3.9}
P_{ij}^{(0)}=p\delta_{ij},\quad {\bf q}^{(0)}=0,
\end{equation}
where $p=nT$ is the hydrostatic pressure. The distribution $f^{(0)}$ is
essentially given by the Sonine approximation (\ref{2.11}) with the cumulant
$c$ given by Eq.\ (\ref{2.12}). 

The analysis to first order in $\epsilon$ is similar to the one woked out
in Ref. \cite{BDKS98} for the free cooling case. We only display here the
final expressions for the fluxes with some details being given
in Appendix \ref{appA}. The final result to first order in the spatial
gradients is 
\begin{equation}
\label{3.13}
P_{ij}^{(1)}=-\eta\left(\nabla_iu_j+\nabla_ju_i-\frac{2}{d}\delta_{ij}
\nabla\cdot{\bf u}\right),
\end{equation}
\begin{equation}
\label{3.14}
{\bf q}^{(1)}=-\kappa \nabla T-\mu \nabla n,
\end{equation} 
where $\eta$ is the shear viscosity, $\kappa$ is the thermal conductivity, 
and $\mu$ is an additional transport coefficient not present in the elastic 
case. In dimensionless form, the transport coefficients are given by  
\begin{equation}
\label{3.15}
\eta^*=\frac{\eta}{\eta_0}=\frac{1}{\nu_{\eta}^*-\zeta^{*}},
\end{equation}
\begin{equation}
\label{3.16}
\kappa^*=\frac{\kappa}{\kappa_0}=
\frac{d-1}{d}\frac{1+c}{\nu_{\kappa}^*-\frac{3}{2}
\zeta^*},
\end{equation}
\begin{equation}
\label{3.17}
\mu^*=\frac{n\mu}{T\kappa_0}=
\frac{d-1}{2d}\frac{c}{\nu_{\mu}^*-\frac{3}{2}
\zeta^{*}}.
\end{equation}
Here, $\eta_0$ and $\kappa_0$ are the elastic values of the shear viscosity 
and thermal conductivity \cite{FK72}, respectively, 
\begin{equation}
\label{3.17.a}
\eta_0=\frac{d+2}{8}\pi^{-(d-1)/2}\Gamma\left(d/2\right)(mT)^{1/2}
\sigma^{-(d-1)},\quad \kappa_0=\frac{d(d+2)}{2(d-1)}\frac{\eta_0}{m},
\end{equation}
$\zeta^*=\zeta^{(0)}/\nu_0$, $\nu_0=p/\eta_0$ is a 
characteristic collision frequency, and $c(\alpha)$ is related to the 
fourth moment of $f^{(0)}$ by 
\begin{equation}
\label{3.18}
c(\alpha)=\frac{8}{d(d+2)}\left[\left(\frac{m}{2T}\right)^2\frac{1}{n}
\int\, d{\bf v} V^4f^{(0)}-\frac{d(d+2)}{4}\right].
\end{equation}
The coefficient $c(\alpha)$ is a measure of the deviation of the reference 
state from that of a gas with elastic collisions. As said above, a good 
estimate of $c(\alpha)$ is given by the Sonine approximation (\ref{2.12}). 
Furthermore, in Eqs.\ (\ref{3.15})--(\ref{3.17}), we have introduced the 
reduced collision frequencies
\begin{equation}
\label{3.19}
\nu_\eta^*=\frac{\int d{\bf v} D_{ij}({\bf V}){\cal L}{\cal C}_{ij}({\bf V})}
{\nu_0\int d{\bf v}D_{ij}({\bf V}){\cal C}_{ij}({\bf V})},
\end{equation}
\begin{equation}
\label{3.20}
\nu_\kappa^*=\frac{\int d{\bf v} {\bf S}({\bf V})\cdot{\cal L}{\sf {\cal 
A}}({\bf V})}
{\nu_0\int d{\bf v}{\bf S}({\bf V})\cdot{\sf {\cal A}}({\bf V})}, \quad 
\nu_\mu^*=\frac{\int d{\bf v} {\bf S}({\bf V})\cdot{\cal L}{\sf {\cal 
B}}({\bf V})}
{\nu_0\int d{\bf v}{\bf S}({\bf V})\cdot{\sf {\cal B}}({\bf V})},
\end{equation}
where 
\begin{equation}
\label{3.21}
D_{ij}({\bf V})=m\left(V_iV_j-\frac{1}{d}\delta_{ij}V^2\right),\quad
{\bf S}({\bf V})=\left(\frac{m}{2}V^2-\frac{d+2}{2}T\right){\bf V}.
\end{equation}

So far, all the results are exact but not explicit because of $\zeta^*$,
$\nu_{\eta}^*$, $\nu_{\kappa}^*$, and $\nu_{\mu}^*$.
To get more explicit expressions for the
dependence of the transport coefficients on $\alpha$ it is convenient to use 
the leading Sonine polynomial approximations for ${\sf {\cal A}}({\bf V})$, 
${\sf {\cal B}}({\bf V})$, ${\cal C}_{ij}({\bf V})$, and $f^{(0)}$. In the 
case of $f^{(0)}$, we take the approximation (\ref{2.11}) while the 
remaining quantities are given by 
\begin{equation}
\label{3.22}
\left( 
\begin{array}{c}
{\sf {\cal A}}({\bf V}) \\ 
{\sf {\cal B}}({\bf V}) \\ 
{\cal C}_{ij}({\bf V}) 
\end{array}
\right) \rightarrow f_{M}(V)\left( 
\begin{array}{c}
c_{T}{\bf S}({\bf V}) \\ 
c_{n}{\bf S}({\bf V}) \\ 
c_{u}D_{ij}({\bf V}) 
\end{array}
\right) ,\hspace{0.3in}f_{M}(V)= n \pi^{-d/2}v_0^{-d}
e^{-\left( V/v_{0}\right) ^{2}} \;. 
\end{equation}
The factor $f_{M}(V)$ occurs since these polynomials are defined relative to
a Gaussian scalar product. The coefficients are the projections of 
${\sf {\cal A}}$, ${\sf {\cal B}}$, and ${\cal C}_{ij}$ along 
${\bf S}({\bf V})$, and $D_{ij}({\bf V})$,  
\begin{equation}
\label{3.23}
\left( 
\begin{array}{c}
c_{T} \\ 
c_{n}
\end{array}
\right) =\frac{2}{d(d+2)}\frac{m}{nT^3} \int d{\bf v} \left( 
\begin{array}{c}
{\sf {\cal A}}({\bf V})\cdot {\bf S}({\bf V}) \\ 
{\sf {\cal B}}({\bf V})\cdot {\bf S}({\bf V})
\end{array}
\right) =\left( 
\begin{array}{c}
-\frac{2}{d+2}\frac{m}{nT^2}\kappa  \\ 
-\frac{2}{d+2}\frac{m}{T^3}\mu  
\end{array}
\right)  \;,
\end{equation}
\begin{equation}
\label{3.24}
c_{u}=\frac{1}{(d+2)(d-1)}\frac{1}{nT^2}
\int d{\bf V}{\cal C}_{ij}({\bf V})D_{ij}(
{\bf V})=-\frac{1}{nT^2}\eta  \;.
\end{equation}
where use has been made of the definitions (\ref{a8}), (\ref{a9}), and 
(\ref{a10}). With these expressions, the cooling rate $\zeta^*$ and the
collision frequencies $\nu_{\eta}^*$, $\nu_{\kappa}^*$, and $\nu_{\mu}^*$
can be explicitly computed. These calculations have been made by Brey and
Cubero \cite{BC01}, with the result
\begin{equation}
\label{3.25}
\zeta^*=\frac{d+2}{4d}(1-\alpha^2)\left(1+\frac{3}{32}c\right),
\end{equation}
\begin{equation}
\label{3.26}
\nu_\eta^*=\frac{3}{4d}\left(1-\alpha+\case{2}{3}d\right)(1+\alpha)
\left(1-\frac{c}{64}\right),
\end{equation}
\begin{equation}
\label{3.27}
\nu_\kappa^*=\nu_\mu^*=
\frac{1+\alpha}{d}\left[\frac{d-1}{2}+\frac{3}{16}(d+8)
(1-\alpha)+\frac{4+5d-3(4-d)\alpha}{1024}c\right],
\end{equation}
where $c$ is given by Eq.\ (\ref{2.12}).

Substitution of the above expressions into Eqs.\ (\ref{3.15})--(\ref{3.17}) 
gives finally the explicit dependence of the transport coefficients on the 
restitution coefficient $\alpha$. In order to gain some insight into the 
behavior of $\eta^*$, $\kappa^*$, and $\mu^*$ it is convenient to consider 
the weak dissipation limit. In this limit, the above coefficients can be 
expanded in powers of the inelasticity parameter $1-\alpha^2$. The leading 
contributions are
\begin{equation}
\label{3.27.1}
\eta^*\approx 1+\frac{64d^2-97d+32}{128d(d-1)}(1-\alpha^2)+\cdots, 
\end{equation}
\begin{equation}
\label{3.27.2}
\kappa^*\approx 1+\frac{56d-191}{128(d-1)}(1-\alpha^2)+\cdots, 
\end{equation}
\begin{equation}
\label{3.27.3}
\mu^*\approx \frac{1}{4(1-d)}(1-\alpha^2)+\cdots. 
\end{equation}

\subsection{Stochastic thermostat}

In the case of the stochastic thermostat (\ref{2.14}), the Boltzmann 
equation becomes 
\begin{equation}
\label{3.29}
\partial_{t}f+{\bf v}\cdot \nabla f
-\frac{1}{2}\frac{T}{m}\zeta \left(\frac{\partial}{\partial {\bf 
V}}\right)^2f=J\left[f,f\right] \;.
\end{equation}
As before, this equation can be solved by means of the 
Chapman-Enskog method. Since the 
procedures to arrive at the expressions of the transport coefficients are 
identical to the ones employed in the case of the Gaussian force, here we 
only quote the final expressions of the transport coefficients.
In dimensionless form, they are 
\begin{equation}
\label{3.31}
\eta^*=\frac{1}{\nu_{\eta}^*},
\end{equation}
\begin{equation}
\label{3.32}
\kappa^*=\frac{d-1}{d}\frac{1+c}{\nu_{\kappa}^*},
\end{equation}
\begin{equation}
\label{3.33}
\mu^*=\frac{d-1}{2d}\frac{c}{\nu_{\mu}^*}.
\end{equation}
Here, in the first Sonine approximation, the expressions of $\zeta^*$, 
$\nu_\eta^*$, and $\nu_\kappa^*$ are also given by Eqs.\ 
(\ref{3.25})--(\ref{3.27}), respectively, with $c(\alpha)$ given by Eq.\ 
(\ref{2.15}). In the quasielastic limit, one has  
\begin{equation}
\label{3.33.1}
\eta^*\approx 1+\frac{32d^2-129d+96}{128d(d-1)}(1-\alpha^2)+\cdots, 
\end{equation}
\begin{equation}
\label{3.33.2}
\kappa^*\approx 1+\frac{8d-287}{128(d-1)}(1-\alpha^2)+\cdots, 
\end{equation}
\begin{equation}
\label{3.33.3}
\mu^*\approx \frac{1}{4(1-d)}(1-\alpha^2)+\cdots. 
\end{equation}

\subsection{Comparison with the free cooling case}

The transport coefficients for a dilute {\em unforced} granular gas
in $d$ dimensions has
been recently obtained by Brey and Cubero \cite{BC01}. These authors  
generalize a previous derivation made from the Boltzmann equation in the 
three dimensional case \cite{BDKS98}. The results are
\begin{equation}
\label{3.34}
\eta^*=\frac{1}{\nu_{\eta}^*-\case{1}{2}\zeta^*},
\end{equation}
\begin{equation}
\label{3.35}
\kappa^*=\frac{d-1}{d}\frac{1+c}{\nu_{\kappa}^*-2\zeta^{*}},
\end{equation}
\begin{equation}
\label{3.36}
\mu^*=\frac{2\zeta^*}{2\nu_{\mu}^*-3\zeta^{*}}
\left[\kappa^*+\frac{(d-1)c}{2d\zeta^*}\right],
\end{equation} 
where $c(\alpha)$ is given by Eq.\ (\ref{2.12}). In the quasielastic limit, 
these coefficients behave as
\begin{equation}
\label{3.37}
\eta^*\approx 1+\frac{48d^2-113d+64}{128d(d-1)}(1-\alpha^2)+\cdots, 
\end{equation}
\begin{equation}
\label{3.38}
\kappa^*\approx 1+\frac{72d-159}{128(d-1)}(1-\alpha^2)+\cdots, 
\end{equation}
\begin{equation}
\label{3.39}
\mu^*\approx \frac{d+1}{4(d-1)}(1-\alpha^2)+\cdots. 
\end{equation}

Comparison between Eqs.\  (\ref{3.27.1})--(\ref{3.27.3}), 
(\ref{3.33.1})--(\ref{3.33.3}), and (\ref{3.37})--(\ref{3.39}) shows that  
the leading order corrections to the elastic values of the transport 
coefficients are clearly different in the free cooling and driven cases. 
This illustrates the fact that the transport properties are affected by the 
thermostat introduced so that the latter does not play a neutral role in the 
problem. In Figs.\ \ref{fig1}, \ref{fig2}, and \ref{fig3} we plot the 
reduced coefficients $\eta^*$, $\kappa^*$, and $\mu^*$, respectively, for 
hard spheres ($d=3$) in the cases of the unforced gas, the Gaussian 
thermostat, and the stochastic thermostat. In general, we observe that the 
predictions obtained in the Gaussian case for the shear viscosity and the
thermal conductivity are closer to those of the
unforced case than to the ones obtained with the stochastic force. This is in 
part motivated by the fact that, in dimensionless form, the results 
of the unforced and Gaussian cases are identical in the uniform problem 
(reference state). In the case of the shear viscosity, all the theories give 
the same trends for $\eta^*$ since this coefficient increases with the 
dissipation. However, at a qualitative level, the influence of dissipation 
on the viscosity in the stochastic case is much less significant than in the 
other two cases. Thus, for instance, for $\alpha=0.8$ (moderate 
dissipation), the shear viscosity of the granular gas $\eta$ has only 
changed about $1\%$ 
with respect to its elastic value when the gas is 
heated by means of the stochastic force. This could justify the use of the 
elastic values of the transport coefficients in the analysis of the 
long-range correlations in a randomly driven granular fluid \cite{NETP99}.
Discrepancies between unforced and the two driven systems
are more important in the case of the thermal conductivity
$\kappa^*$, as Fig.\ \ref{fig2} shows. In addition, while in the free cooling  
and Gaussian cases $\kappa^*$ increases with $\alpha$, the opposite happens 
in the stochastic case. Finally, Fig.\ \ref{fig3} shows the dependence of 
the coefficient $\mu^*$ on $\alpha$. We observe that $\mu^*\simeq 0$ 
for both thermostats, while this coefficient is clearly different from 
zero in the free cooling problem. This is basically due to the fact that 
$\mu^*\propto c(\alpha)$ in the Gaussian and stochastic cases, Eqs.\ 
(\ref{3.17}) and (\ref{3.33}), and so it  
vanishes exactly if one takes the Maxwellian approximation (which is known 
to give a very accurate description) for the reference state. This means 
that, for practical purposes, one can neglect the contribution to the heat 
flux coming from the term proportional to the density gradient in the heated 
gas case: ${\bf q}^{(1)}\to -\kappa \nabla T$. As is apparent from Fig.\ 
\ref{fig3}, this cannot be assumed in the unforced case since $\mu^*$ is 
clearly different from zero even in the quasielastic limit.

\section{Comparison with Monte Carlo simulations}
\label{sec4}

As has been discussed above, the practical evaluation of
the transport coefficients requires the truncation of
an expansion for the solutions
of the integral equations in Sonine polynomials. In the case of
elastic collisions, the leading order truncation is known to be a very good 
approximation \cite{FK72}. A natural question is whether the above degree of 
accuracy is also maintained in the inelastic case. To answer this 
question, one has to resort to numerical solutions of the Boltzmann 
equation, such as those obtained from the Direct Simulation Monte Carlo 
(DSMC) method \cite{B94}. Although this method was originally devised for 
molecular fluids, its extension to deal with inelastic collisions is
straightforward\cite{MS00,F00}.

Recently, the shear viscosity of a low-density granular gas of freely evolving
hard spheres has been determined from the DSMC method \cite{BRC99}.
The results show a very good agreement with the predictions based on the 
Boltzmann equation in the first Sonine approximation, Eq.\ (\ref{3.34}). This
experiment consists in preparing an initial inhomogeneous
nonequilibrium state corresponding to a transverse shear wave, and then 
analyzing its subsequent evolution in time. The 
shear wave decays exponentially with a time scale inversely proportional to 
the viscosity. An alternative route to measuring the shear viscosity
consists of preparing a state of uniform shear flow using
Lees-Edwards boundary conditions\cite{LE72}.
Macroscopically, this state is characterized by constant
density $n$, a uniform temperature $T$, and a linear velocity profile 
$a=\partial u_x/\partial y\equiv\text{const}$. In a molecular fluid, unless 
a termostatting force is introduced, the temperature increases
in time due to viscous heating. The corresponding energy balance equation 
can be used to determine the shear viscosity for sufficiently long times 
\cite{NO79}. In a granular fluid, the relationship between
the temperature and the shear viscosity is not simple since
there is a competition between viscous
heating and collisional cooling \cite{MGSB99,TTMGSD01}.
However, if external forces of the form (\ref{2.13}) and
(\ref{2.14}) that exactly
compensate for the collisional energy loss are introduced, the viscous heating 
effect is still able to heat the system. Under these conditions, the 
Boltzmann equation to be solved is
\begin{equation}
\label{4.1}
\partial_tf-aV_y\frac{\partial}{\partial V_x}f+
\frac{1}{2}\zeta\frac{\partial}{\partial {\bf V}}
\cdot\left({\bf V}f\right)
=J\left[f,f\right]  
\end{equation}
in the case of the Gaussian thermostat, and
\begin{equation}
\label{4.2}
\partial_tf-aV_y\frac{\partial}{\partial V_x}f
-\frac{1}{2}\frac{T}{m}\zeta \left(\frac{\partial}{\partial {\bf 
V}}\right)^2f=J\left[f,f\right] \;
\end{equation}
in the case of the stochastic thermostat.

Equations (\ref{4.1}) and (\ref{4.2}) have been numerically solved
by means of the DSMC method for a three-dimensional system ($d=3$).
At given values of the shear rate $a$ and the restitution
coefficient $\alpha$, we start from a local equilibrium state and monitor 
the time evolution of $a^*=a/\nu_0$ and $P_{xy}^*=P_{xy}/p$. Here, 
$\nu_0=p/\eta_0\propto T^{1/2}$ is an effective collision frequency,
$\eta_0$ is the elastic shear viscosity. The simulations
show that, after a transient regime, the ratio $-P_{xy}^*/a^*$ reaches a 
constant value (independent of the shear rate), which can be identified as
the shear viscosity in the linear hydrodynamic regime. Details
of the simulation for dense fluids will be published elsewhere 
\cite{MGSD02}, and here only
compare the Monte Carlo simulations for the shear viscosity with the Sonine 
approximations given by Eqs.\ (\ref{3.15}) and (\ref{3.31}).

In our simulations we have typically taken $10^5$ particles and have averaged over
5 replicas. Since the thermal velocity $v_0$ is not constant in the transient regime,
we have taken a time-dependent time step given by $0.01 \ell /v_0(t)$, where
$\ell=(\sqrt{2}\pi n \sigma^2)^{-1}$ is the mean free path.
The simulation results are shown in Fig.\ \ref{fig1}. In
general, we observe a very good agreement between the predictions
of the Chapman-Enskog theory in the first Sonine approximation
and the simulation data. This agreement is
similar to the one previously found in the free cooling case \cite{BRC99}.
At a quantitative level, we see that the discrepancies between theory 
and simulation tend to increase as the dissipation increases, although these
differences are quite small (less than $3\%$). 
As a final conclusion, it is important to remark that the agreement extends over a wide range of values
of the restitution coefficient ($\alpha\geq 0.6$), indicating the
reliability of the Sonine approximation for describing granular flows
beyond the quasielastic domain.

\section{Summary and discussion}
\label{sec5}

In this paper we have addressed the derivation of the hydrodynamic equations 
of a granular gas from the Boltzmann kinetic theory. The system is 
heated by the action of ``thermostatting'' external forces which 
exactly compensate for cooling effects associated with the inelasticity 
of collisions. Two different types of thermostats have been considered: (a) an 
``anti-drag'' force proportional to the particle peculiar velocity (Gaussian 
force), and (b) a stochastic force, which gives frequent kicks to each 
particle between collisions. The introduction of these thermostats has the 
advantage of avoiding the intrinsic time dependence of the homogeneous 
cooling state (unforced gas), but at the price of introducing unknown new 
effects induced by the external forcing. While in the homogeneous problem 
the results obtained with and without a Gaussian thermostat are completely 
equivalent (when one scales the particle velocity 
with the thermal velocity), this equivalence does not hold
in the stochastic case and
the non-Maxwellian properties of the distribution function (cumulants) are 
different from those obtained in the unforced case. Here, our goal has been to 
assess the influence of thermostats on the transport 
properties of the gas.

The transport processes considered are those for a fluid with small spatial 
gradients of the hydrodynamic fields. For this reason, the Boltzmann 
equation has been solved perturbatively using an adaptation of the
Chapman-Enskog method recently proposed for inelastic
collisions \cite{BDKS98,GD99}. By using similar procedures
as those made in the free cooling case\cite{BDKS98}, we calculate the
distribution function to first order in the gradients.
Its use in the functionals for the pressure tensor and heat flux 
provides a representation of these as linear combinations of the gradients. 
The corresponding coefficients in these expressions are the shear viscosity 
$\eta$ (defined in Eq.\ (\ref{3.13})), the thermal conductivity $\kappa$ and 
the new coefficient $\mu$ (both coefficients defined in Eq.\ (\ref{3.14})). 
These transport coefficients are in general functions of the restitution 
coefficient $\alpha$. Their expressions are given by Eqs.\ 
(\ref{3.15})--(\ref{3.17}) in the case of the Gaussian thermostat and Eqs.\ 
(\ref{3.31})--(\ref{3.33}) in the case of the stochastic thermostat. 
A practical evaluation of these coefficients is possible by means of a 
Sonine polynomial approximation and the derivation and approximate results 
are not limited to weak inelasticity.

The dependence of $\eta$, $\kappa$, and $\mu$ on $\alpha$ has been 
illustrated in the case of hard spheres ($d=3$). As Figs.\ \ref{fig1}--\ref{fig3}
show, the thermostats affect the transport properties
since the discrepancies between the driven and free cooling results are quite 
significant. Although not widely recognized, the above conclusion 
illustrates the fact that generally the inclusion of an external force
depending on the state of the system changes the apparent transport
coefficients. This has been demonstrated as well for uniform shear flow with 
elastic collisions where thermostats are used to produce a steady state. In 
that case, the Navier-Stokes shear viscosity is unchanged, but nonlinear 
rheological properties are affected by the thermostat \cite{DSBR86}. In the 
context of granular fluids, the effects already occur at the Navier-Stokes 
order. Notice that the above conclusion only affects to this type of external
forcing mechanisms (termostats), since driving the system by shaking, vibration,
and even the action of a weak external field (such as the gravity field) 
does not modify the transport coefficients of the gas.
Concerning the influence of dissipation on transport, we observe that
in general the deviation from the functional form for elastic collisions is 
more significant in the unforced gas than in the driven cases. In particular, 
the coefficient $\mu$ (which is zero in the elastic limit) is clearly 
different from zero in the unforced gas (for instance, $\mu^*\simeq 0.27$ 
at $\alpha=0.8$) while is negligible in the Gaussian and stochastic cases
(for instance, at $\alpha=0.5$, $\mu^*\simeq0.064$ for the Gaussian force while
$\mu^*\simeq0.012$ for the stochastic force).

To check the accuracy of the Sonine approximation, we have numerically solved
the Boltzmann equation by means of the Direct Simulation Monte Carlo 
(DSMC) method\cite{B94} for a granular gas under uniform shear flow. To 
control inelastic cooling, a thermostat force is introduced in the 
system. In these conditions, the (apparent) Navier-Stokes shear viscosity 
can be measured directly in the long time limit just as 
for the case of elastic collisions. This simulation method has been 
recently proved \cite{MGSD02} to be an efficient way of measuring the shear 
viscosity of a moderately dense granular gas. The comparison carried out here in 
the low-density regime shows that the Chapman-Enskog results
in the first Sonine approximation exhibit an
excellent agreement with the simulation data, even for moderate  
dissipation (say, for instance $\alpha=0.7$). 
These results indicate clearly the 
reliability of the quantitative predictions for transport 
coefficients from the Chapman-Enskog method with small spatial gradients
but including strong dissipation.

As said before, a study of long-range correlations in a granular system 
fluidized by the random stochastic force (\ref{2.12}) has been recently 
made \cite{NETP99}. In order to analyze the decay of fluctuations, it 
was assumed that the transport coefficients are equal to the corresponding 
quantities given by the elastic Enskog theory \cite{FK72}.
According to the results obtained in this paper for a heated granular gas
in the low-density regime, although this assumption
can be considered as a quite good approximation 
for the shear viscosity $\eta$ 
and the coefficient $\mu$, this is not true for the thermal conductivity 
since $\kappa^*$ clearly differs from 1, even for small inelasticity. In
this context, it would be interesting to reexamine the conclusions
obtained in Ref.\ \onlinecite{NETP99} when the true Enskog
transport coefficients are considered. As said in the Introduction, the
Enskog equation for a gas of inelastic hard spheres ($d=3$) in the
absence of thermostats has been recently solved \cite{GD99} up to Navier-Stokes
order. Taking into account these results \cite{GD99}, the extension of the
calculations carried out in this paper to higher densities can be easily 
made. The final expressions of the corresponding Enskog transport 
coefficients in the three dimensional case are displayed in Appendix 
\ref{appB}. We plan to extend these expressions for a $d$ dimensional 
granular fluid in the near future.

\acknowledgments 
V.G. acknowledges partial support from the Ministerio de Ciencia y 
Tecnolog\'{\i}a (Spain) through Grant No. BFM2001-0718.

\appendix
\section{Chapman-Enskog expansion}
\label{appA}

In the case of the Gaussian thermostat, the velocity distribution $f^{(1)}$ 
obeys the kinetic equation 
\begin{equation}
\label{a1}
\left({\cal L}+\frac{1}{2}\zeta^{(0)}
\frac{\partial}{\partial {\bf V}}\cdot {\bf 
V}\right)f^{(1)}+\frac{1}{2}\zeta^{(1)}
\frac{\partial}{\partial {\bf V}}\cdot \left({\bf V}f^{(0)}\right)
=-\left(D_t^{(1)}+{\bf V}\cdot \nabla\right)f^{(0)},
\end{equation}
with $D_t^{(1)}=\partial_t^{(1)}+{\bf u}\cdot \nabla$, ${\cal L}$ is the 
linear operator given by
\begin{equation}
\label{3.11}
{\cal L}f^{(1)}=-\left(J[f^{(0)},f^{(1)}]+J[f^{(1)},f^{(0)}]\right),
\end{equation}
and $\zeta^{(1)}$ is a linear functional of $f^{(1)}$ defined as
\begin{equation}
\label{a1.bis}
\zeta^{(1)}=-\frac{1}{dnT}\int\, d{\bf v} 
mV^2\left(J[f^{(0)},f^{(1)}]+J[f^{(1)},f^{(0)}]\right).
\end{equation}

The macroscopic balance equations to first order in the gradients are
\begin{equation}
\label{a2}
D_t^{(1)}n=-n\nabla\cdot {\bf u},\quad 
D_t^{(1)}u_i=-(mn)^{-1}\nabla_ip, \quad 
D_t^{(1)}T=-\frac{2T}{d}\nabla\cdot {\bf u}.
\end{equation}
Use of these in Eq.\ (\ref{a1}) yields
\begin{equation}
\label{a3}
\left({\cal L}+\frac{1}{2}\zeta^{(0)}
\frac{\partial}{\partial {\bf V}}\cdot {\bf 
V}\right)f^{(1)}+\frac{1}{2}\zeta^{(1)}
\frac{\partial}{\partial {\bf V}}\cdot \left({\bf V}f^{(0)}\right)
={\bf A}\cdot 
\nabla \ln T+{\bf B}\cdot \nabla \ln n+C_{ij}\nabla_iu_j,
\end{equation}
where the expressions of ${\bf A}$, ${\bf B}$, and $C_{ij}$ are the same
as those obtained in the Appendix A of Ref.\ \cite{BDKS98}, i.e., 
\begin{equation}
\label{a4}
{\bf A}({\bf V})=\frac{1}{2}{\bf V}\frac{\partial}{\partial {\bf 
V}}\cdot\left({\bf V}f^{(0)}\right)-\frac{T}{m}
\frac{\partial}{\partial {\bf V}}f^{(0)},
\end{equation}
\begin{equation}
\label{a5}
{\bf B}({\bf V})=-{\bf V}f^{(0)}-\frac{T}{m}
\frac{\partial}{\partial {\bf V}}f^{(0)},
\end{equation}
\begin{equation}
\label{a6}
C_{ij}({\bf V})=\frac{\partial}{\partial 
V_i}\left(V_jf^{(0)}\right)-\frac{1}{d}
\delta_{ij}\frac{\partial}{\partial {\bf V}}\cdot \left({\bf 
V}f^{(0)}\right).
\end{equation}
The fact that $C_{ij}$ is traceless implies that the scalar $\zeta^{(1)}=0$ 
by symmetry. This is special to the low density Boltzmann equation, since at 
higher densities \cite{GD99} there is a contribution to $f^{(1)}$ 
proportional to $\nabla \cdot {\bf u}$ leading to a nonzero value of 
$\zeta^{(1)}$. This can be seen in Appendix \ref{appB}.
Comparison with the kinetic equation obeying $f_1^{(1)}$ in the
unforced case (Eq.\ (A5) of Ref.\ \cite{BDKS98}) shows that both kinetic
equations only differ in the terms $\partial_t^{(0)} f_1^{(1)}$ (unforced
description) and $(\zeta^{(0)}/2)(\partial/\partial {\bf V}) \cdot {\bf V}
f_1^{(1)}$ (driven case). These terms give different contributions
to the transport coefficients. The solution to Eq.\ (\ref{a3}) is of the form
\begin{equation}
\label{3.12}
f^{(1)}={\sf {\cal A}}\cdot \nabla \ln T+{\sf {\cal B}}\cdot 
\nabla \ln n+{\cal C}_{ij} \nabla_iu_j.
\end{equation}
Substituting Eq.\ (\ref{3.12}) into Eq.\ (\ref{a1}) and identifying
coefficients of independent gradients, one gets
the following three integral equations determining
the unknowns ${\sf {\cal A}}({\bf V})$, ${\sf {\cal B}}({\bf V})$, and 
${\cal C}_{ij}({\bf V})$:
\begin{equation}
\label{a7}
\left({\cal L}+\frac{1}{2}\zeta^{(0)}
\frac{\partial}{\partial {\bf V}}\cdot {\bf 
V}\right)
\left(
\begin{array}{c}
{\sf {\cal A}}\\
{\sf {\cal B}}\\
{\cal C}_{ij}
\end{array}
\right)=
\left(
\begin{array}{c}
{\bf A}\\
{\bf B}\\
C_{ij}
\end{array}
\right).
\end{equation}
 
The expressions (\ref{3.15}), (\ref{3.16}), and (\ref{3.17}) for the 
transport coefficients follow directly from the integral equations 
(\ref{a7}). The transport coefficients are defined as \cite{BC01} 
\begin{equation}
\label{a8}
\eta=-\frac{1}{(d-1)(d+2)}\int d{\bf V} D_{ij}({\bf V}) {\cal C}_{ij}({\bf 
V}),
\end{equation}
\begin{equation}
\label{a9}
\kappa=-\frac{1}{dT}\int\, d{\bf v} {\bf S}({\bf V})\cdot {\sf {\cal 
A}}({\bf V}),
\end{equation}
\begin{equation}
\label{a10}
\mu=-\frac{1}{dn}\int\, d{\bf v} {\bf S}({\bf V})\cdot {\sf {\cal 
B}}({\bf V}).
\end{equation}
The thermal conductivity $\kappa$ and the coefficient $\mu$ can be easily
obtained by multiplying the two first equations of (\ref{a7})
by ${\bf S}({\bf V})$ and integrating over the velocity. The result is 
\begin{equation}
\label{a11}
\left(\nu_{\kappa}-\case{3}{2}\zeta^{(0)}\right)\kappa=
\frac{d+2}{2}\frac{nT}{m}(1+c),
\end{equation}
\begin{equation}
\label{a12}
\left(\nu_{\mu}-\case{3}{2}\zeta^{(0)}\right)\mu=
\frac{d+2}{4}\frac{T^2}{m}c.
\end{equation}
Upon deriving these expressions, use has been made of the relation
\begin{equation}
\label{a13}
\frac{1}{2}\frac{\partial}{\partial {\bf V}}\cdot \left({\bf V}f^{(0)}\right)
=-T\partial_Tf^{(0)}
\end{equation}
which follows from the temperature dependence of $f^{(0)}$.
The shear viscosity $\eta$ can be obtained in a similar way by
multiplying the third equation of (\ref{a7}) by $D_{ij}({\bf V})$:
\begin{equation}
\label{a14}
\left(\nu_{\eta}-\zeta^{(0)}\right)\eta=nT.
\end{equation}
Equations (\ref{a11}), (\ref{a12}), and (\ref{a14}) lead directly
to expressions (\ref{3.15})--(\ref{3.17}) appearing in the text.

The analysis in the case of the stochastic thermostat (\ref{2.13}) is
similar to that made above for the Gaussian one. To first order in gradients,
one has 
\begin{equation}
\label{a18}
\left[{\cal L}-\frac{1}{2}T\frac{\zeta^{(0)}}{m}
\left(\frac{\partial}{\partial {\bf V}}\right)^2\right]
f^{(1)}
={\bf A}\cdot 
\nabla \ln T+{\bf B}\cdot \nabla \ln n+C_{ij}\nabla_iu_j.
\end{equation}
Here, the velocity dependence on the right side of Eq.\ (\ref{a18}) is given 
by Eqs.\ (\ref{a4})--(\ref{a6}) and we have taken into account that 
$\zeta^{(1)}=0$. The solution to Eq.\ (\ref{a18}) is of the form 
(\ref{3.12}), where now the corresponding integral equations are
\begin{equation}
\label{a19}
\left[{\cal L}-\frac{1}{2}T\frac{\zeta^{(0)}}{m}
\left(\frac{\partial}{\partial {\bf V}}\right)^2\right]
\left(
\begin{array}{c}
{\sf {\cal A}}\\
{\sf {\cal B}}\\
{\cal C}_{ij}
\end{array}
\right)=
\left(
\begin{array}{c}
{\bf A}\\
{\bf B}\\
C_{ij}
\end{array}
\right).
\end{equation}
Now, we multiply Eq.\ (\ref{a19}) by ${\bf S}({\bf V})$ and $D_{ij}({\bf 
V})$ and integrate over the velocity to get the expressions of 
$\kappa$, $\mu$, and $\eta$:
\begin{equation}
\label{a20}
\nu_{\kappa}\kappa=\frac{d+2}{2}\frac{nT}{m}(1+c),
\end{equation}
\begin{equation}
\label{a21}
\nu_{\mu}\mu=\frac{d+2}{4}\frac{T^2}{m}c,
\end{equation}
\begin{equation}
\label{a22}
\nu_{\eta}\eta=nT.
\end{equation}
Here, we have made use of the results
\begin{equation}
\label{a23}
\int d{\bf v}\, {\bf S}({\bf V})\cdot \left(\frac{\partial}{\partial {\bf 
V}}\right)^2{\sf {\cal A}}=0, 
\end{equation} 
\begin{equation}
\label{a24}
\int d{\bf v}\, {\bf S}({\bf V})\cdot \left(\frac{\partial}{\partial {\bf 
V}}\right)^2{\sf {\cal B}}=0, 
\end{equation} 
\begin{equation}
\label{a25}
\int d{\bf v}\, D_{ij}({\bf V})\left(\frac{\partial}{\partial {\bf 
V}}\right)^2{\cal C}_{ij}=0,
\end{equation} 
which follow directly from the solubility conditions (\ref{3.3}) and 
(\ref{3.4}).

\section{Results for a dense fluid of hard spheres}
\label{appB}

In this Appendix, we display the results derived for inelastic hard spheres
($d=3$) in the framework of the Enskog equation when the gas is heated
by the Gaussian and the stochastic thermostats.
To first order in the gradients, the momentum and heat fluxes are 
\begin{equation}
\label{b1}
P_{ij}^{(1)}=-\eta\left(\nabla_iu_j+\nabla_ju_i-\frac{2}{3}\delta_{ij}
\nabla\cdot{\bf u}\right)-\gamma \delta_{ij}\nabla\cdot {\bf u},
\end{equation}
\begin{equation}
\label{b2}
{\bf q}^{(1)}=-\kappa \nabla T-\mu \nabla n,
\end{equation} 
where $\gamma$ is the bulk viscosity coefficient which vanishes in the
low density limit. In a compact form, the transport coefficients can be
written as
\begin{equation}
\label{b3}
\eta=\eta^{k}\left[1+\frac{2\pi n^*\chi(1+\alpha)}{15}\right]+\frac{3}{5}
\gamma,
\end{equation}
\begin{equation}
\label{b4}
\gamma=\frac{2}{9}\sqrt{\pi m T}n^* n\chi \sigma (1+\alpha)
\left(1-\frac{c}{32}\right),
\end{equation}
\begin{equation}
\label{b5}
\kappa=\kappa^{k}\left[1+\frac{\pi n^*\chi(1+\alpha)}{5}\right]+
\frac{1}{3}\sqrt{\frac{\pi T}{m}}n^* \chi n \sigma
\left(1+\frac{7c}{32}\right),
\end{equation}
\begin{equation}
\label{b6}
\mu=\mu^{k}\left[1+\frac{\pi n^*\chi(1+\alpha)}{5}\right],
\end{equation}
where $\chi$ is the pair correlation function at contact, $n^*=n\sigma^3$ is
a reduced density, and the superscript $k$ denotes the contributions from the
kinetic parts of the fluxes\cite{GD99}. These kinetic parts are given by 
\begin{equation}
\label{b7}
\eta^{k}=nT\left(\nu_{\eta}-b_{\eta}\zeta^{(0)}\right)^{-1}\left[1-\frac{1}{15}
(1+\alpha)(1-3\alpha)\pi n^*\chi\right],
\end{equation}
\begin{eqnarray}
\label{b8}
\kappa^k&=&\frac{5}{2}\frac{nT}{m}
\left(\nu_{\kappa}-b_{\kappa}\zeta^{(0)}\right)^{-1}\left\{1+
\left[1+c\left(1+\frac{1+\alpha}{6}\pi n^*\chi\right)\right]\right.
\nonumber\\
& & \left. +\frac{1}{10}\pi n^*\chi(1+\alpha)^2\left[2\alpha-1
+\left(\frac{1}{2}(1+\alpha)-\frac{5}{3(1+\alpha)}\right)c\right]\right\},
\end{eqnarray}
\begin{eqnarray}
\label{b9}
\mu^k&=&\frac{T}{n}
\left(\nu_{\mu}-b_{\mu}\zeta^{(0)}\right)^{-1}\left\{\frac{5}{4}\frac{nT}{m}
\left(1+\frac{1+\alpha}{3}\pi n^*\chi\right)\left[1+n\partial_n\ln \left(
1+\frac{1+\alpha}{3}\pi n^*\chi\right)\right]c \right. \nonumber\\
& & \left. -\frac{nT}{2m}\pi n^*\chi(1+\alpha)\left(1+\frac{1}{2}
n\partial_n\ln \chi \right)\left[\alpha(1-\alpha)+\frac{1}{4}\left(
\alpha(1-\alpha)+\frac{4}{3}\right)c\right]\right\}.
\end{eqnarray}
In these equations, $b_{\eta}=-1$, and $b_{\kappa}=b_{\mu}=-\case{3}{2}$ in the
case of the Gaussian thermostat while $b_{\eta}=b_{\kappa}=b_{\mu}=0$ for the
stochastic thermostat.

Up to first order in gradients, the cooling rate $\zeta$ is given by
\begin{equation}
\label{b10}
\zeta\to \frac{5}{12}\nu_0\chi(1-\alpha^2)\left(1+\frac{3c}{32}\right)
+\left[-\frac{1}{3}\pi n^* \chi+\frac{5}{32}\nu_0
\left(1+\frac{3c}{64}\right)
\chi \beta\right] (1-\alpha^2)\nabla\cdot {\bf u},
\end{equation}
where the expression of the quantity $\beta$ depends on the thermostat used.
In the Gaussian case, one has 
\begin{equation}
\label{b11}
\beta=\pi n^* \chi\frac{(2/45)\lambda-(c/6)(1+\alpha)
(5-3\alpha)}{\nu_{\gamma}-2\zeta^{(0)}-(5c/64)(1-\alpha^2)
\left(1+\frac{3c}{64}\right)\chi},
\end{equation}
while in the stochastic case one gets
\begin{equation}
\label{b12}
\beta=\pi n^* \chi\frac{(2/45)\lambda-(c/3)(1+\alpha)}{\nu_{\gamma}}.
\end{equation}
Here, we have introduced the quantities\cite{GD99}
\begin{equation}
\label{b13}
\lambda=\frac{3}{8}(1-\alpha^2)\left(5\alpha^2+4\alpha-1+\frac{c}{12}
\frac{159\alpha+3\alpha^2-19\alpha-15\alpha^3}{1-\alpha}\right),
\end{equation}
\begin{equation}
\label{b14}
\nu_{\gamma}=\frac{1+\alpha}{48}\nu_0\chi\left[128-96\alpha+15\alpha^2
-15\alpha^3+\frac{c}{64}\left(15\alpha^3-15\alpha^2+498\alpha-434\right)
\right].
\end{equation}

To get the explicit dependence of the transport coefficients and the
cooling rate on the reduced density $n^*$, one can take for instance the
Carnahan-Starling approximation for $\chi$ given by
\begin{equation}
\label{b15}
\chi=\frac{1-\case{1}{12}\pi n^*}{1-\case{1}{6}\pi n^*}
\end{equation}
It is easy to check that all results of Appendix \ref{appB} reduce to
those presented in the text for the low-density limit ($n^*=0$).

\begin{figure}
\caption{Reduced shear viscosity $\eta^*=\eta/\eta_0$  
as a function of the restitution coefficient $\alpha$ for a three-dimensional 
system. The solid lines refer to the theoretical expressions derived 
in the unforced case (a), in the Gaussian case (b), and in the stochastic 
case (c). The symbols are the results obtained from the Direct Simulation 
Monte Carlo method in the Gaussian (circles) and stochastic (triangles) cases.  
\label{fig1}}
\end{figure}
\begin{figure}
\caption{Reduced thermal conductivity $\kappa^*=\kappa/\kappa_0$  
as a function of the restitution coefficient $\alpha$ for a three-dimensional 
system in the unforced case (a), in the Gaussian case (b), and in the stochastic 
case (c). 
\label{fig2}}
\end{figure}
\begin{figure}
\caption{Reduced coefficient $\mu^*=n\mu/T\kappa_0$  
as a function of the restitution coefficient $\alpha$ for a three-dimensional 
system in the unforced case (a), in the Gaussian case (b), and in the stochastic 
case (c). 
\label{fig3}}
\end{figure}


\begin{references}

\bibitem{FK72}J. Ferziger and H. Kaper, Mathematical Theory of 
Transport Processes in Gases, North-Holland, Amsterdam, 1972.


\bibitem{BDKS98}J. J. Brey, J. W. Dufty, C. S. Kim, A. Santos, Phys. 
Rev. E 58 (1998) 4638.

\bibitem{BC01}J. J. Brey, D. Cubero, in Granular Gases, edited by 
T. P\"oschel and S. Luding, Lecture Notes in Physics, Springer Verlag, 
Berlin, 2001, p. 59. 


\bibitem{GD99}V. Garz\'o, J. W. Dufty, Phys. Rev. E 59 (1999) 5895. 
 

\bibitem{YHCMW02}X. Yiang, C. Huan, D. Candela, R. W. Mair,
R. L. Walsworth, Phys. Rev. Lett. 88 (2002) 044301.

\bibitem{D01} See for instance, the review of J. W. Dufty, cond-mat/0108444.

\bibitem{WM96}D. R. M. Williams, F. C. McKintosh, Phys. Rev. E 54
(1996) R9.

\bibitem{varios}A. Puglisi, V. Loreto, U. M. B. Marconi, A. Petri, A. 
Vulpiani, Phys. Rev. Lett. 81 (1998) 3848; Phys. Rev. E 59 (1999) 5582;
R. Cafiero, S. Luding, H. J. Herrmann, Phys. Rev. Lett. 84 (2000) 6014;
G. Peng, T. Ohta, Phys. Rev. E 58 (1998) 4737; S. J. Moon,
M. D. Shattuck, J. B. Swift, Phys. Rev. E 64 (2001) 031303. 

\bibitem{NE98}T. P. C. van Noije, M. H. Ernst, Granular Matter 1 (1998) 57.

\bibitem{NETP99}T. P. C. van Noije, M. H. Ernst, E. Trizac, I. 
Pagonabarraga, Phys. Rev. E 59 (1999) 4326.

\bibitem{PTNE02}I. Pagonabarraga, E. Trizac, T. P. C. van Noije, M. H. 
Ernst, Phys. Rev. E 65 (2002) 011303.

\bibitem{MS00}J. M. Montanero and A. Santos, Granular Matter
2 (2000) 53.


\bibitem{EM90}D. J. Evans, G. P. Morriss, Statistical Mechanics of
Nonequilibrium Liquids, Academic Press, London, 1990.


\bibitem{BSSS99}C. Bizon, M. D. Shattuck, J. B. Swift, H. L. Swinney,
Phys. Rev. E 60 (1999) 4340. 

\bibitem{GS95}A. Goldshtein, M. Shapiro, J. Fluid Mech. 282 (1995) 75. 

\bibitem{BDS97}J. J. Brey, J. W. Dufty, A. Santos, J. Stat. Phys. 87
(1997) 1051.

\bibitem{BRC96}J. J. Brey, M. J. Ruiz-Montero, D. Cubero,
Phys. Rev. E 54 (1996) 3664.


\bibitem{F00}A. Frezzotti, Physica A 278 (2000) 161.

\bibitem{B94}G. Bird, Molecular Gas Dynamics and the Direct Simulation 
of Gas Flows, Clarendon, Oxford, 1994.

\bibitem{BRC99}J. J. Brey, M. J. Ruiz-Montero, D. Cubero, Europhys. 
Lett. 48 (1999) 359.

\bibitem{LE72}A. W. Lees, S. F. Edwards, J. Phys. C 5 (1972) 1921.
 
\bibitem{NO79}T. Naitoh, S. Ono, J. Chem. Phys. 70 (1979) 4515.

\bibitem{MGSB99}J. M. Montanero, V. Garz\'o, A. Santos,
J. J. Brey, J. Fluid Mech. 389 (1999) 391.

\bibitem{TTMGSD01}M. Tij, E. Tahiri, J. M. Montanero, V. Garz\'o,
A. Santos, J. W. Dufty, J. Stat. Phys. 103 (2001) 1035.


\bibitem{MGSD02}J. M. Montanero, V. Garz\'o, A. Santos, J. W. Dufty, 
{\em Shear viscosity for a moderately dense granular gas} (unpublished). 

\bibitem{DSBR86}J. W. Dufty, A. Santos, J. J. Brey, R. F. Rodr\'{\i}guez,
Phys. Rev. A 33 (1986) 459; J. J. Dufty, J. J. Brey, A. Santos,
in Molecular Dynamics Simulation of Statistical-Mechanical Systems,
edited by G. Ciccotti and W. Hoover, Elsevier, Amsterdam, 1986, p. 294;
V. Garz\'o, A. Santos, J. J. Brey, Physica A 163 (1990) 651. 


\end{references}
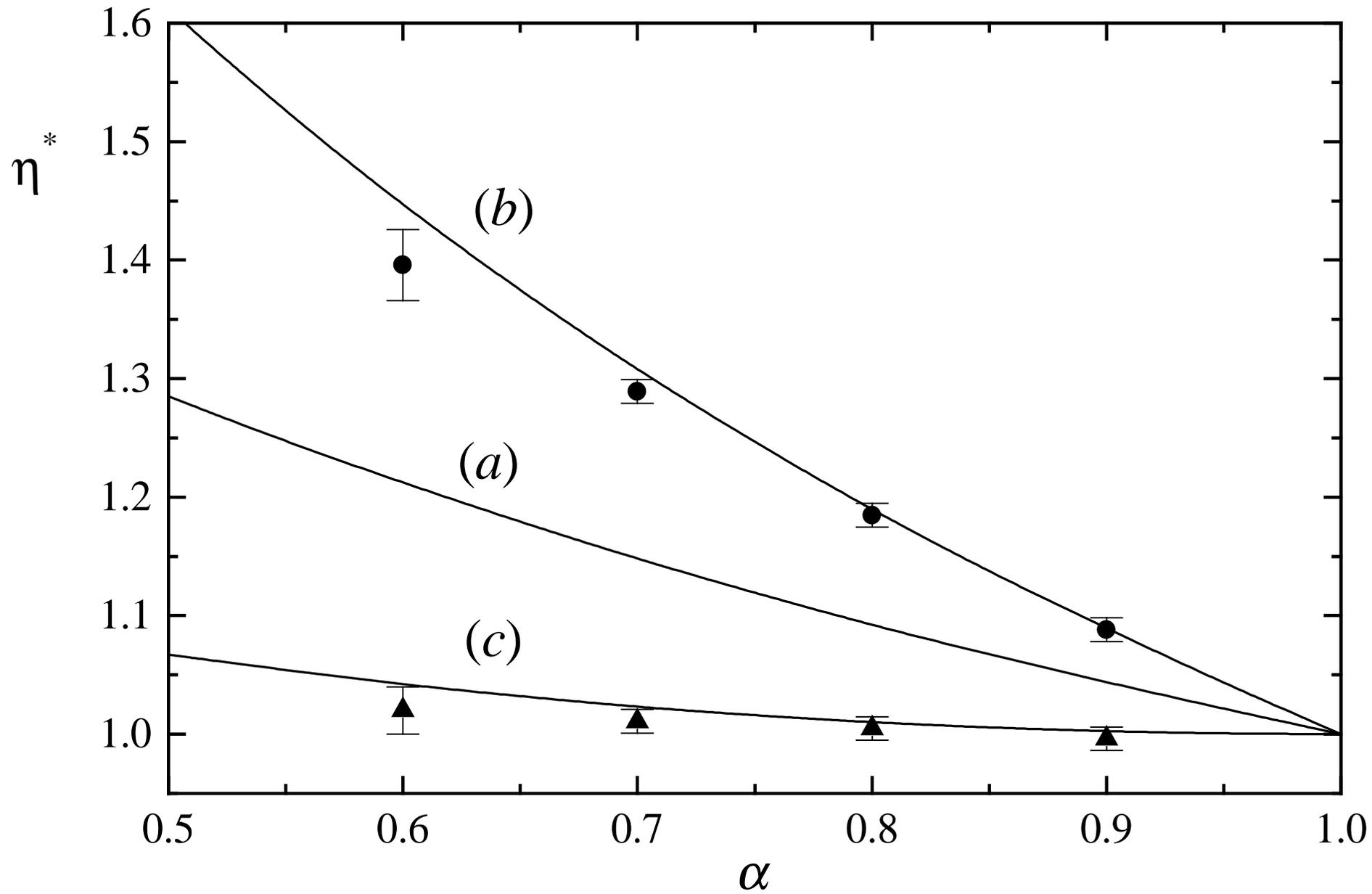
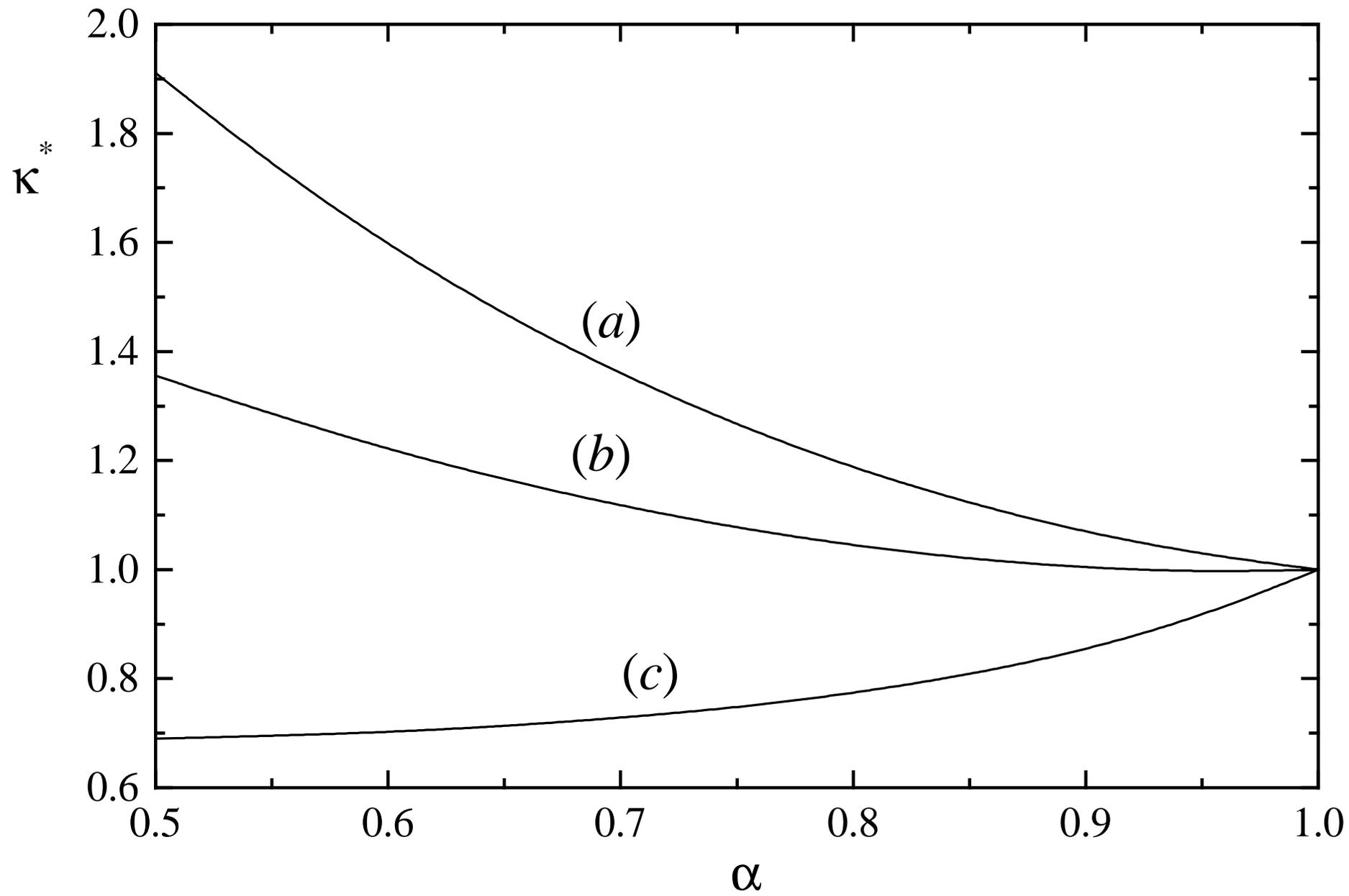
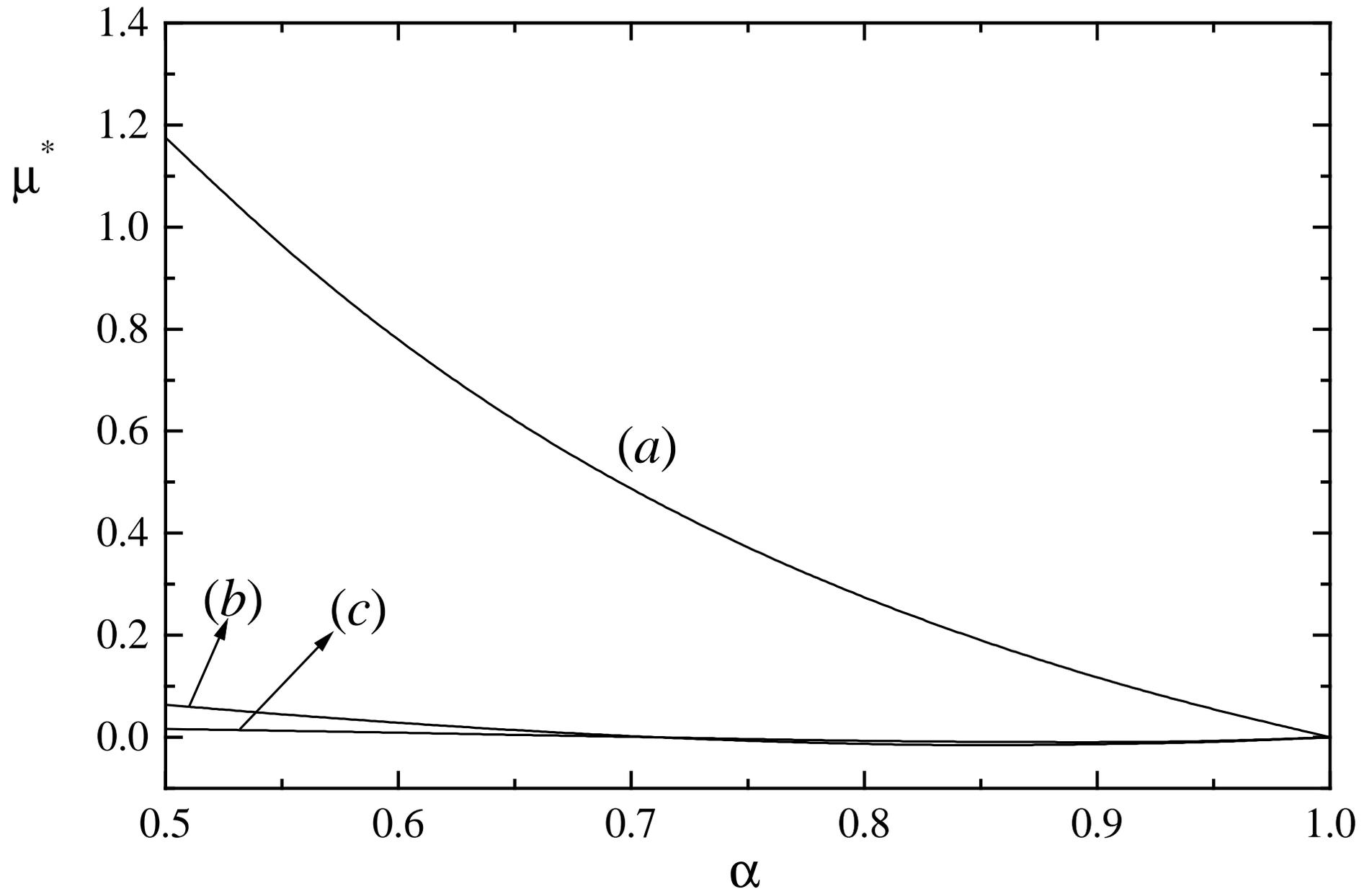
\end{document}